# Trends in large scale structure observations and the likelihood of early reionization

Andrew R. Liddle[1] and David H. Lyth[2]
[1] *Astronomy Centre, University of Sussex, Falmer, Brighton BN1 9QH, U. K.*
[2] *School of Physics and Materials, University of Lancaster, Lancaster LA1 4YB, U. K.*

**ABSTRACT**
With the imminent promise of constraints on the epoch of reionization from observations of microwave background anisotropies, the question of whether or not the standard Cold Dark Matter (CDM) model permits early reionization has been subjected to detailed investigation by various authors, with the conclusion that reionization may occur at quite high redshift. However, it is widely accepted that this model is excluded, as when normalised to the COBE observations it possesses excessive galaxy clustering on scales below tens of megaparsecs. We examine the trends of observations, first in a fairly model independent way, and second by considering variants on the standard CDM model introduced to resolve the observational conflicts. We conclude that the epoch of reionization favoured by the observational data is typically considerably later than the standard CDM model suggests, and amongst models which may fit the observational data only the introduction of a cosmological constant leads to a reionization redshift close to that of standard CDM.

**Key words:** cosmology: theory – dark matter – diffuse radiation.

## 1 INTRODUCTION

One of the most important tests that a model of large scale structure must pass is the Gunn-Peterson test, which indicates that the universe is nearly completely reionized by a redshift of five or greater (Gunn & Peterson 1965; Schneider et al. 1989). Until recently, there has been little prospect of imposing an upper limit on the ionization redshift, but that is now promised via measurements of microwave background anisotropies on degree scales and less; if reionization is sufficiently early then the expected 'Doppler' peak on scales of around a degree may be sharply suppressed due to the photons scattering off the reionized electrons at lower redshifts (Sugiyama et al. 1993). The ever increasing catalogue of anisotropy experiments already hints strongly that the peak is present at around the correct height for models based on adiabatic perturbations without reionization, arguing against very early reionization (Scott et al. 1994).

Reionization provides a fascinating link between different regions of the power spectrum of density perturbations (Couchman & Rees 1986), since it is normally assumed that it is induced by early formation of massive stars which produce an ionizing flux. The first objects to form are usually assumed to be galaxies of mass around $10^6 M_\odot$, corresponding to the mass enclosed in a comoving region of diameter around 0.01 Mpc, but their effect on the microwave background alters the interpretation of anisotropies which originate from perturbations on scales of hundreds of megaparsecs (Vittorio & Silk 1984; Bond & Efstathiou 1984). Reionization will also raise the Jeans' mass which may influence the formation of structures somewhat larger than the ones which induce the reionization (Couchman & Rees 1986).

Recently, two detailed papers have appeared, one by Tegmark, Silk & Blanchard (1994) (hereafter referred to as TSB) and one by Fukugita & Kawasaki (1994) (hereafter FK), which examine the question of reionization in the standard Cold Dark Matter (CDM) model of structure formation. The standard CDM model comprises a spatially flat universe dominated by cold dark matter, containing baryons with density consistent with standard nucleosynthesis (though many authors assume zero baryon density), and with an adiabatic, gaussian, scale-invariant spectrum of density perturbations. We shall define precisely what we take as the standard CDM model later. [TSB, and in addition a follow-up paper by Tegmark & Silk (1994), also investigated several other models. However, they did not normalise their spectra to the COBE observations; the final result is extremely sensitive to the normalisation.] Although the approaches of TSB and FK were rather different, the former concentrating on estimation and the latter on numerical simulation, very similar results were obtained. The conclusion was that with reasonable parameter values, the CDM model will reionize early, at a redshift around 35; optimistic parameter choices obviously lead to yet earlier reionization.

This is easily early enough to have an important impact on degree scale anisotropies (Sugiyama et al. 1993).

However, it is widely accepted that standard CDM is excluded by large scale structure observations, because when normalised to COBE it possesses excessive power on scales below some tens of megaparsecs. Since the epoch of reionization is very sensitive to the amount of power in the spectrum on very short scales, the suspicion is therefore that observations favour substantially later reionization. It is our aim in this paper to take advantage of the detailed work by TSB and FK to study the implications of large scale structure observations for the reionization epoch, at first making as little reference to specific models of structure formation as we can, and then discussing features specific to the different ways in which one can conceive of extending standard CDM to satisfy the observations. In cold dark matter dominated models with critical density, the analysis is particularly simple; other models feature more complicated evolution of density perturbations at moderate redshift which must be taken into account. Important aspects of the results we present are that throughout we shall normalise the spectra to COBE, using the recent results of Bunn, Scott & White (1994), and that we shall incorporate the effect of baryons into the transfer functions we use.

## 2 THEORY AND OBSERVATIONS

### 2.1 Standard CDM and its variants

All of the models considered here assume that large scale structure originates as an adiabatic gaussian density perturbation, whose spectrum at horizon entry is more or less scale independent with any scale dependence parametrised as a power $k^{n-1}$ of the comoving wavenumber $k$. In order to have any chance of agreeing with observations these models require non-baryonic cold dark matter.

In general, such models are specified by the spectral index $n$, the Hubble parameter $H$ whose present value is parametrised as $100h\,{\rm km\,s^{-1} Mpc^{-1}}$, the nature and density of the non-baryonic dark matter, and the value of the cosmological constant. The density of baryons as a fraction of the critical density, $\Omega_B$, is determined by nucleosynthesis to be $\Omega_B h^2 = 0.013 \pm 0.002$. The standard CDM model has $n = 1$, $h = 0.5$, pure cold dark matter, total density $\Omega_{\rm mat} = 1$ and no cosmological constant. Normalised to the COBE observation it gives too much power on smaller scales, so a number of variants of it are currently under consideration. These are conveniently characterised by modifying just one assumption of the standard CDM model, though it is important to recognise that nature may have chosen to modify at least two of them. The options under active investigation at present are

(i) The tilted[*] model (Bond 1992; Liddle et al. 1992; Cen et al. 1992) which has spectral index $n < 1$.

(ii) The low Hubble constant model (Bartlett et al. 1994) which has $h < 0.5$.

(iii) The Mixed Dark Matter (MDM) model (Bonometto & Valdarnini 1984; Shafi & Stecker 1984; Davis et al. 1992a), which replaces some of the CDM by hot dark matter (normally assumed to be a massive neutrino) with density $\Omega_\nu$.

(iv) The cosmological constant model which replaces some of the CDM by a cosmological constant corresponding to a vacuum energy density $\Omega_{\rm vac}$.

(v) The open universe model which simply throws away some of the CDM.

### 2.2 Strategy for calculating the reionization redshift

Following TSB, we estimate the reionization redshift in three stages.

First we need the fraction $f$ of baryons that has to be bound into stars in order to provide sufficient ionizing flux that the reionized fraction is unity. This is uncertain by orders of magnitude. However, the joy of examining reionization in the context of models based on gaussian random fields is that this fraction is an exponentially sensitive function of the amplitude of perturbations and hence of redshift, resulting in an uncertainty of only a modest factor in calculating the latter.

Throughout we shall use the estimates of $f$ derived by TSB. They write $f$ as a product of several factors each of which is estimated separately, and conclude that the 'Middle-of-the-Road' estimate for the collapsed fraction required to induce complete reionization is $f \simeq 8 \times 10^{-3}$. The optimistic and pessimistic estimates (where in their terminology 'optimistic' refers to earlier reionization) are given as $f \simeq 4 \times 10^{-5}$ and $f \simeq 0.8$ respectively. Note that the pessimistic estimate is very pessimistic indeed, being a product of pessimistic estimates for four separate factors (likewise for the optimistic estimate). The largest individual uncertainty is the fraction of hydrogen burned within a short time of star formation. Once reionization has occurred, the ionizing flux which continues to be emitted is enough to prevent subsequent recombination (Fukugita & Kawasaki 1994).

The next step is to suppose that this fraction $f$ is given by $f(>M_{\rm min}, z_{\rm ion})$, where $f(>M, z)$ is the fraction[†] of the mass of the universe that is bound into objects with mass bigger than $M$ at redshift $z$, and $M_{\rm min}$ is the minimum galaxy mass at the redshift of reionization. Two criteria are of relevance for estimating this minimum mass. The baryonic Jeans' mass governs pressure support against collapse, and the mass must be large enough that the cooling time, which governs dissipation, is no greater than the dynamical time (Blanchard et al. 1992). These two criteria are broadly similar, and the normal assumption is that $M_{\rm min}$ is around $10^6 M_\odot$, with leeway of around an order of magnitude in either direction.

The final step is to estimate $f(>M, z)$ within a given model using the Press-Schechter approximation (Press & Schechter 1974). This states that $f(>M, z)$ is equal to twice the fraction of space in which the density contrast, smoothed on the scale $M$, exceeds a threshold $\delta_c$. Since the probability distribution of the filtered density contrast is gaussian, this gives

$$f(>M, z) = \text{erfc}\left(\frac{\delta_c}{\sqrt{2}\,\sigma(M,z)}\right), \qquad (1)$$

where $\sigma(M, z)$ is the dispersion of the density contrast smoothed on a scale $M$ at redshift $z$, and 'erfc' is the complementary error function.

---

[*] In an inflationary context, the introduction of tilt is accompanied by the possibility of a gravitational wave contribution to COBE (Davis et al. 1992b; Liddle & Lyth 1992). This will always make reionization more recent, and we shall neglect this possibility here.

[†] In the hope of avoiding confusion, we have eschewed the usual notation $\Omega(>M,z)$, as $\Omega$ conventionally gives the fraction of a critical density, whereas for reionization we are interested in the fraction of the total matter density, regardless of its value.

To complete the definition of the Press-Schechter approximation one has to specify the filter function used to perform the smoothing, and the threshold $\delta_c$.[‡] We shall use a top-hat filter, and take the threshold value $\delta_c = 1.7$ which is motivated by a spherical collapse model. A gaussian filter is also commonly employed. For a given $M$ the gaussian-smoothed dispersion is higher than the top-hat-smoothed dispersion by a significant scale-dependent factor (Liddle & Lyth 1993). This means that the equivalent value of $\delta_c$ is significantly lower, the equivalence being of course only approximate because the relative factor is scale-dependent.

The Press-Schechter approximation is expected to be reasonable provided that the mass fraction, or equivalently the dispersion, is significantly less than unity. In other words, it is expected to be valid in the linear regime, where the filtered density contrast is evolving according to linear cosmological perturbation theory. $N$-body simulations can check its validity (as long as the mass fraction is not too small), and also provide an empirical 'best fit' value for $\delta_c$. Results vary (Efstathiou & Rees 1988; Brainerd & Villumsen 1992; White et al. 1993; Lacey & Cole 1994; Ma & Bertschinger 1994), but such simulations typically suggest that the approximation is at least roughly correct, with $\delta_c$ lying within twenty percent or so of the theoretically motivated 1.7. As one expects, the values of $\delta_c$ suggested for Gaussian smoothing are typically somewhat lower than those suggested for top hat smoothing.

With pure CDM and critical density, the evolution of $\sigma(R, z)$ in the linear regime is simply

$$\sigma(R, z) = \sigma(R, 0)/(1 + z), \qquad (2)$$

where $\sigma(R, 0)$ is the linearly evolved quantity at the present epoch. Substituting this expression into (1) gives

$$1 + z_{\text{ion}} = \frac{\sqrt{2}\,\sigma(M_{\min}, 0)}{\delta_c}\text{erfc}^{-1}(f). \qquad (3)$$

This is the formula we use for the optimistic and Middle-of-the-Road estimates of $f$. With the pessimistic estimate $f \simeq 0.8$, most of the mass is supposed to have collapsed before reionization occurs. In that case the quasi-linear theory underlying the Press-Schechter approximation ceases to be reliable, and we replace it by the crude estimate that a significant fraction of mass collapses when $\sigma(M_{\min}, z) \sim 1$, leading to

$$1 + z_{\text{ion}} = \sigma(M_{\min}, 0). \qquad (4)$$

### 2.3 The transfer function

The top-hat filter of comoving radius $R$ is given by

$$W(kR) = \frac{\sin kR}{(kR)^3} - \frac{\cos kR}{(kR)^2}, \qquad (5)$$

where $k$ is the comoving wavenumber. The relation between mass and radius for the top-hat filter is

$$M = 1.16 \times 10^{12}\,\Omega_{\text{mat}}\,h^{-1}\left(\frac{R}{h^{-1}\text{Mpc}}\right)^3 M_\odot. \qquad (6)$$

Following the notation of Liddle & Lyth (1993), the present dispersion is

$$\sigma^2(R, 0) = 9H_0^{-4}\int_0^\infty k^4 T^2(k)\,\delta_H^2(k)\,W^2(kR)\,\frac{dk}{k}. \qquad (7)$$

The transfer function $T(k)$ is normalised to unity as $k \to 0$. The initial spectrum $\delta_H^2(k)$ is defined here to be independent of $k$ for a flat ($n = 1$) spectrum; for a tilted spectrum one has $\delta_H^2(k) \propto k^{n-1}$. The usual quantity $P(k)$ is proportional to $k\delta_H^2(k)$.

The transfer function for a given version of the CDM model can be calculated by numerical integration of the relevant evolution equations. Unfortunately, published calculations are not completely adequate for our purpose and we shall have to do a certain amount of improvisation. For the moment let us consider the case where all the dark matter is cold.

Both TSB and FK take the transfer function from Bardeen et al. (1986) (henceforth BBKS) without further comment. However, there are two important questions. Firstly, calculation of transfer functions to very small scales is troublesome because of the difficult integration over many oscillations of the baryon fluid; most published transfer functions are only accurate down to $1h^{-1}$ Mpc at best (see the discussion in Liddle & Lyth (1993)). Although the BBKS transfer function has the correct asymptotic form including the logarithmic correction to the large $k$ behaviour, they supply no indication of the accuracy of its amplitude on scales as short as we are interested in. Secondly, the BBKS transfer function is for zero baryon density (for which the usual $\Omega h$ scaling is exact); however, baryons will reduce the short scale power, and even at $8h^{-1}$ Mpc this effect is known to be around 10% for $h = 0.5$.

There is little we can do about the former problem, but we can at least attempt to allow for the latter via an empirical scaling of the transfer function with baryon density as publicised by Peacock & Dodds (1994). On intermediate scales this brings the BBKS transfer function into good agreement with calculations including baryons (at least for the range of $\Omega_B$ we are interested in); that the BBKS transfer function then possesses the right asymptotic form, unlike other parametrisations, may to some extent address the question of accuracy at short scales too. We therefore take the transfer function as

$$T_{\text{CDM}}(q) = \frac{\ln(1 + 2.34q)}{2.34q} \times$$

---

[‡] Neither TSB nor FK are very specific about the type of smoothing. TSB use a top-hat filter but quote some lower values of $\delta_c$ which are at least in part lower because the papers cited used a gaussian filter. FK use a gaussian filter but take $\delta_c = 1.69$ which is normally motivated by top-hat collapse.

4     *A. R. Liddle and D. H. Lyth*

$$\left[1 + 3.89q + (14.1q)^2 + (5.46q)^3 + (6.71q)^4\right]^{-1/4}, \quad (8)$$

where $q = k/[\Omega_{\mathrm{mat}} h^2 \exp(-2\Omega_B)\,\mathrm{Mpc}^{-1}]$.

## 2.4 Observations

The above estimate of the reionization redshift needs as input the value of $\sigma(R, 0)$ on the scale $R \simeq 0.01$ Mpc. Observations which have been used to constrain large scale structure theories do not reach down to such small scales, but they do cover four orders of magnitude from roughly the size of the observable universe down to about one megaparsec, with fairly tight constraints available on various scales. Most of the data can be summarised as a value of $\sigma(R, 0)$ which allows it to be presented in a single plot, Figure 1. A full discussion of the data requires a separate paper (Liddle et al. 1994) [see also Lyth & Liddle (1994)], and here we shall only summarise it with particular emphasis on the crucial COBE normalisation. Our summary shall concentrate on $\Omega_{\mathrm{mat}} = 1$; a full generalisation to arbitrary $\Omega_{\mathrm{mat}}$ will be given elsewhere though we shall mention some aspects later.

Over the desired range, the dispersion varies by several orders of magnitude while we are interested in differences of a factor two. Consequently, we shall plot all the data as its ratio to the prediction of standard CDM; since standard CDM certainly fits all the data to within a factor two or so (though not within the actual errors), this greatly clarifies the graphical presentation.

In Figure 1 the error bars indicate something like a 1-$\sigma$ range, but the upper and lower limits indicate rather firm results.

(i) **The COBE measurement of microwave anisotropies.** The most accurate measurements of the microwave background anisotropy presently available are those of the COBE satellite. The analysis and interpretation of these measurements remains an evolving field. A very elegant analysis was recently carried out by Górski et al. (1994). Using the Sachs-Wolfe approximation for the $C_l$ (the expected value of the $l$-th anisotropy multipole), they found that a good fit is obtained with a spectral index $n$ anywhere in the range 0.6 to 1.4, and that all fits gave essentially the same value for $C_9$. For $n = 1$ they concluded that the expected quadrupole $Q_{\mathrm{rms-PS}} = (19.9 \pm 1.6)\mu$K. Very recently Bunn et al. (1994) have done an independent fit of the Sachs-Wolfe approximation, finding the value $Q_{\mathrm{rms-PS}} = (21.1 \pm 1.6)\mu$K which is slightly but not significantly higher. However, they go on to fit to a full calculation of the predicted anisotropies, including the tail of the Doppler Peak. More or less independently of the nature of the dark matter and the values of $h$ and $\Omega_B$, they find

$$Q_{\mathrm{rms-PS}}(n) = (19.9 \pm 1.5) \exp[0.69(1-n)]\ \mu\mathrm{K}, \quad (9)$$

and this is the normalisation we shall adopt. The relation between $\delta_H$ and $Q_{\mathrm{rms-PS}}$ can be derived analytically for power law spectra (Liddle & Lyth 1993) and for $n = 1$ is $\delta_H = 2.26 \times 10^{-5}(Q_{\mathrm{rms-PS}}/19.9\mu\mathrm{K})$. We shall schematically represent the COBE point as being at $4000 h^{-1}$Mpc, though when discussing specific models they shall be normalised using the Bunn et al. result directly.

(ii) **The galaxy correlation function.** This is very naturally given in terms of $\sigma(R, 0)$. Data are now sufficiently copious to provide constraints over more than an order of magnitude in $R$. We take the values given by the compilation of data sets by Peacock & Dodds (1994), converted to $\sigma(R, 0)$ according to their prescription. In Figure 1 they are represented by a band, whose edges correspond to 1-sigma errors *excluding* overall normalisation. The larger error bars at each end of the band indicate the uncertainty in the overall normalisation, which allows one to shift the entire data set up or down while preserving its shape. The data is scaled for the best fit bias parameter with $\Omega_{\mathrm{mat}}$ held at unity, $b_I = 0.85^{+0.35}_{-0.25}$. This is determined from redshift distortions and nonlinear effects in the spectrum (Peacock & Dodds 1994). However, the following data more stringently constrain the overall normalisation.

(iii) **Peculiar velocity flows.** The bulk flow around us smoothed on a scale $40 h^{-1}$ Mpc is quite well measured (Bertschinger et al. 1990). The velocities are actually sensitive to longer scales in the dispersion $\sigma(R, 0)$, and in practice this measurement samples $\sigma(R, 0)$ on a scale of around $90 h^{-1}$ Mpc (Lyth & Liddle 1994). Unfortunately, as it is a single measurement the 'cosmic variance' is high, dominating the observational errors. An analysis using a comparison of peculiar velocity flows with IRAS galaxy distributions (Dekel et al. 1993) is more useful, giving $b_I = 0.7^{+0.6}_{-0.2}$ at 95% confidence level. Using the upper limit to normalise the Peacock and Dodds data gives the limit $\sigma(8 h^{-1}\,\mathrm{Mpc}, 0) > 0.6$ shown in Figure 1.

(iv) **Galaxy cluster abundance.** This directly measures the dispersion on a scale around $8 h^{-1}$ Mpc, using either the Press-Schechter approximation or numerical simulations. From the calculations of White, Efstathiou & Frenk (1993), we deduce that if the virialised mass of Abell clusters with richness class > 1 is less than $1.5 h^{-1} \times 10^{15} M_\odot$, then $\sigma(8 h^{-1}\mathrm{Mpc}, 0) < 1.0$. This mass is three times as big as the one estimated from temperature and virial velocity measurements, though we note that it might be consistent with gravitational lensing measurements. In any case, it seems reasonable to regard the above bound as rather firm, and it is shown in Figure 1.

(v) **Early object formation.** The spectrum must contain enough power on short scales to reproduce the observations of quasar abundances (Efstathiou & Rees 1988) and the fractional density in damped Lyman alpha systems (Mo & Miralde-Escude 1994; Kauffmann & Charlot 1994). This provides lower limits on $\sigma(R, z)$, where $z$ is the relevant redshift; results can then be scaled to $z = 0$. The lower limits we show assume purely cold dark matter but are not much different in viable MDM models. They should be rather firm as they correspond to very loose interpretations of the measurements (Lyth & Liddle 1994; Liddle et al. 1994).

For our final table of results, it is useful to charac-terise the fit to the data by two parameters, which indicate how well COBE normalised models fit the galaxy correlation function. One is the overall normalisation, which we characterise by $\sigma(8h^{-1}{\rm Mpc}, 0)$. As discussed its allowed range is

$$\sigma(8h^{-1}{\rm Mpc}, 0) = 0.6 \text{ to } 1.0, \tag{10}$$

if $\Omega_{\rm mat} = 1$. The other parameter we need is some measure of the shape. In models without a hot component, we follow Peacock & Dodds (1994) in defining the shape parameter by

$$\Gamma = \Omega h \exp(-2\Omega_B) - 0.32\left(\frac{1}{n} - 1\right). \tag{11}$$

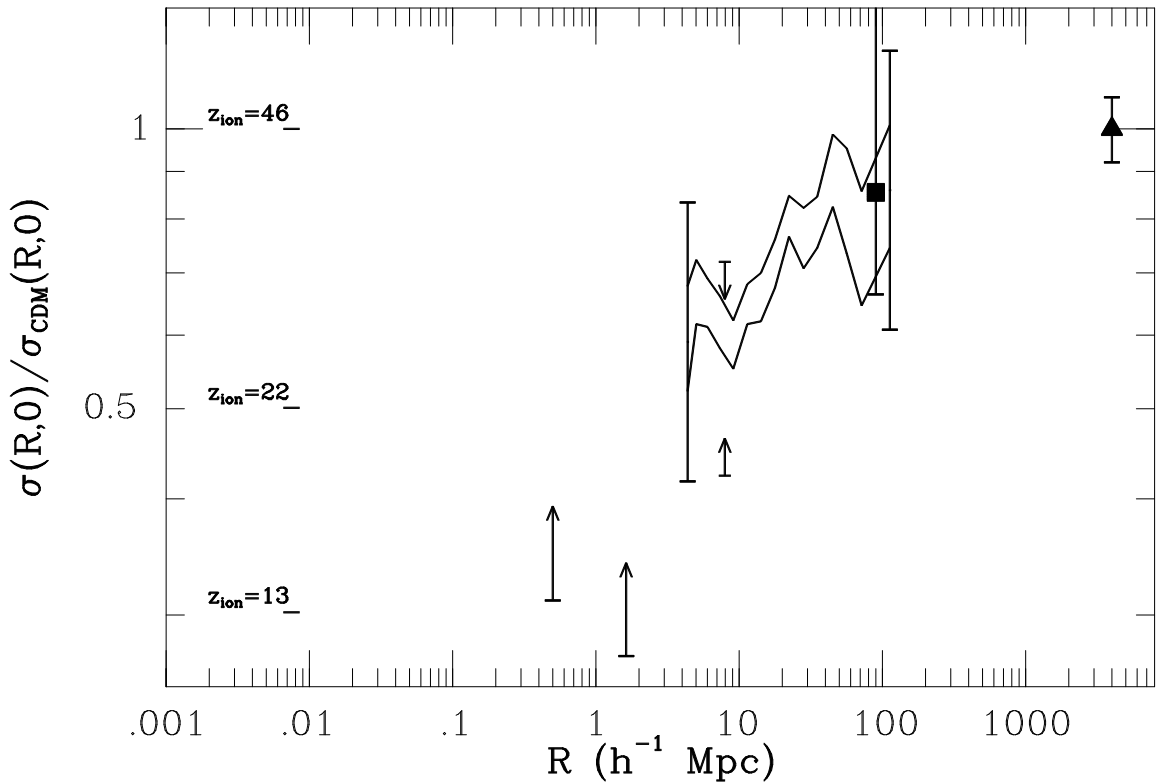

**Figure 1.** Observational values of the present linearly evolved dispersion of the smoothed density contrast. These values apply to critical density CDM and MDM models, but not to low density models. They are plotted as a fraction of the predicted value for the standard CDM model. From the right: the solid triangle represents COBE; the solid square is from the bulk velocity flow; the band indicates the 1-sigma allowed region of the galaxy correlation points according to Peacock & Dodds (1994) (15 points are plotted and the jaggedness is noise) while the error bars at each end of the band indicate the uncertainty in bias from this method; the upper and lower limits at $8h^{-1}$Mpc, coming from cluster abundance and bulk flows respectively, are limits on the normalisation at high confidence which are much stricter than those from the galaxy correlation function alone; the other lower limits are from the quasar abundance (right) and from the amount of gas in damped Lyman alpha systems (left). Finally, the notches on the left indicate reionization redshifts for the standard CDM model (top value), and modifications of it which invoke tilt and/or a low Hubble constant to reduce the small scale power. The estimates of $z_{\rm ion}$ are positioned at the scale corresponding to $M = 10^6 M_\odot$, because the reionization redshift depends only on $\sigma(R,0)$ on that scale. The aim is to extrapolate through the observations to the notches from which the reionization redshift can be directly read off. For MDM models the observational data are almost unchanged, but the reionization redshifts must be rescaled as indicated in the text. For open and cosmological constant models this Figure does not apply at all; they must be treated separately.

In MDM models we estimate an effective $\Gamma$ from the slope of the dispersion at $25h^{-1}$Mpc. At the 2-sigma level the range of $\Gamma$ that fits the data of Peacock & Dodds (1994) is

$$\Gamma = 0.22 \text{ to } 0.29, \tag{12}$$

(see also Kofman et al. (1993)).

## 3 THE EPOCH OF REIONIZATION

In this section we calculate the reionization redshift, first in the standard CDM model, which is ruled out by observation, and then in variants of it which may be consistent with observation. The results shall be displayed in Table 1.

### 3.1 Standard CDM

In the standard CDM model the reionization redshift is given by (3) and (4). Using the COBE normalisation, one obtains $\sigma(M_{\min}, 0) = 33.9, 29.1$ and $24.6$ for $M_{\min} = 10^5 M_\odot$, $10^6 M_\odot$ and $10^7 M_\odot$ respectively. If we take the Middle-of-the-Road estimate for the collapse fraction $f$ required to reionize, we get $z_{\rm ion} = 53, 46$ and $38$ for the three values of $M_{\min}$. On the other hand, keeping $M_{\min}$ fixed, the optimistic and pessimistic estimates for $f$ multiply these figures by 1.5 and 0.6 respectively. As this latter is the larger uncertainty (and also dominates the uncertainty in $\delta_c$), from now on we shall fix $M_{\rm min}$ to be $10^6 M_\odot$. Accounting for our different approach regarding the normalisation of the spectrum and the incorporation of baryons, these results agree with TSB, unsurprisingly as we have followed their estimations. The principle reason for the somewhat higher redshift is the higher normalisation to COBE.

From Figure 1 it is clear that standard CDM is ruled out by observation, so we go on to consider variants of it, retaining always the COBE normalisation. In each case we want to choose the additional free parameter to give a reasonable fit to the slope and magnitude of the galaxy correlation data.

### 3.2 Tilted and low Hubble constant models

For the 'tilted' and 'low Hubble constant' variants of the standard CDM model, perturbation growth is just as in standard CDM, and so (3) and (4) remain valid. Consequently, $1 + z_{\rm ion}$ is just reduced in proportional to $\sigma(M_{\rm min}, 0)$. The aim in altering these parameters is to try and fit the large scale structure observations; extrapolating the data shown in Figure 1 through to the notches, which have been placed at the scale corresponding to $M_{\rm min} = 10^6 M_\odot$, gives the reionization redshift. Without worrying about particular parameter values, with the strong downward trend enforced by the galaxy correlation data it seems hard to imagine that these variants can give a reionization epoch to be much above $z = 20$ in these variants, even with these data points raised collectively upwards as much as allowed.

In Table 1 we give the normalisation and shape parameter for the representative choices $h \simeq 0.3$ and $n \simeq 0.7$. Either of them does a fair job of satisfying the data and they both give reionization around $z_{\rm ion} = 10$.

### 3.3 Mixed Dark Matter models

In an MDM model, a fraction $\Omega_\nu$ of the matter is in the form of hot dark matter. Its free-streaming leads to a scale dependent suppression in the growth of perturbations, but for the short scales relevant to reionization the effect is simply a slowed growth rate across all relevant epochs, with the hot component playing no part in the gravitational instability (Davis et al. 1992a). The growth rate during matter domination is

$$\sigma_{\rm MDM}(M, z) \propto (1+z)^{-\alpha}, \qquad (13)$$

with

$$\alpha = \left(\sqrt{25 - 24\Omega_\nu} - 1\right)/4. \qquad (14)$$

Consequently, if we know the value of $\sigma(10^6 M_\odot, 0)$, it is simple to track it back to determine the reionization redshift.

Unfortunately, there do not exist transfer functions for MDM which are accurate, or even have the right asymptotic form, for the scales we are interested in. Instead, we follow Davis et al. (1992a) and use a sudden transition approximation. This assumes that in an MDM model the perturbation growth up to the end of the radiation era can be taken as the same as CDM; the transition to matter domination is then taken as instantaneous and the suppressed growth law above is assumed to take effect immediately. If accurate transfer functions are obtained in the future, it will be interesting to compare them with this approximation.

A subtlety to be taken into account is that in the standard incarnation of the MDM model, where the hot component is a massive tau neutrino, this neutrino becomes non-relativistic before matter domination, which then occurs earlier than in the usual case where the tau neutrino acts as radiation. The usual epoch $z_{\rm eq} \simeq 24000\Omega_{\rm mat} h^2$ is replaced by $z_{\rm eq} \simeq 27000\Omega_{\rm mat} h^2$, since each neutrino species contributes 12% of the present radiation density.

We shall quote results for two values of the hot component density. The originally favoured $\Omega_\nu = 0.3$ (giving $\alpha = 0.8$) has lost favour due to its apparent inability to explain the amount of gas in damped Lyman alpha systems (Mo & Miralde-Escude 1994; Kauffmann & Charlot 1994; Ma & Bertschinger 1994), and so we shall also quote for $\Omega_\nu = 0.15$ (giving $\alpha = 0.91$). In Table 1 we give the values of $\sigma(8h^{-1}{\rm Mpc}, 0)$ and $\Gamma$ for these two choices, calculating the former from the transfer functions of Schaefer and Shafi (1994) (which are accurate on these scales), and estimating an equivalent value for the latter from the slope of $\sigma_{\rm MDM}(R, 0)$ at $R = 25h^{-1}{\rm Mpc}$. Taking into account the 8% uncertainty in the COBE normalisation, both choices give acceptable values for $\sigma(8h^{-1}{\rm Mpc}, 0)$, and the values of $\Gamma$ are also reasonable though an intermediate choice would do better.

Most of the data in Figure 1 applies at redshift zero, and the correction to the object abundances is small for reasonable $\Omega_\nu$. Consequently, Figure 1 still indicates the trends in the data. However, the reionization values given by the notches must be rescaled, with $1 + z_{\rm ion}$ replaced by $(1 + z_{\rm ion})^{1/\alpha}$. Alternatively, the approximation above gives an analytic formula for the central reionization redshift

$$1 + z_{\rm ion} = 47^{1/\alpha}(27000h^2)^{1-1/\alpha}. \qquad (15)$$

For $\Omega_\nu = 0.3$, the central redshift estimate is $z_{\rm ion} = 13$, while for $\Omega_\nu = 0.15$ it is $z_{\rm ion} = 27$. The growth suppression is playing some role in making these values larger than they would otherwise have been, though particularly in the case of the latter reionization is earlier through the actual normalisation of the power spectrum being rather high. The slower perturbation growth also results in the spread in redshifts from the uncertainty in $f$ being larger than in other cases.

### 3.4 Cosmological constant and open universes

Finally we come to the cosmological constant and open models, where the present matter density $\Omega_{\rm mat}$ is less than one. In these models the COBE normalisation is amended and all observational data need reassessed, so Figure 1 cannot be used at all. We content ourselves by examining specific parameter values, which we choose to be $\Omega_{\rm mat} \simeq 0.3$ and $h \simeq 0.8$. These do as well as any in minimizing the disagreement with observation (but note that for the open case the present age of the universe would be only 10 Gyrs, while with a cosmological constant it is 12 Gyrs).

Before a redshift of at most a few, $\Omega_{\rm mat}$ is close to unity and the growth of perturbations is as in the critical density case. As a result the predicted reionization redshift is the same for a given choice of the spectrum $\delta_H^2(k)$ of the primordial density perturbation. Afterwards the growth slows down, so that by the present the density contrast is suppressed by a factor $g$ relative to the critical density case.

**Table 1.** Estimates of the reionization epoch $z_{\rm ion}$, assuming a minimum collapse mass of $10^6 M_\odot$. The uncertainty in this mass is dominated by the uncertainty in the collapsed baryon fraction required to induce reionization. For information we include the predicted values of the parameters $\sigma(8h^{-1}\,{\rm Mpc}, 0)$ and $\Gamma$, which measure the normalisation and slope of the galaxy correlation function. From observation the target range of $\Gamma$ is 0.22 to 0.29. For critical density models the target range of $\sigma(8h^{-1}\,{\rm Mpc}, 0)$ is 0.6 to 1.0, but for the low density models in the last block it is 0.9 to 1.4.

| Model | $h$ | $n$ | $\Omega_{\rm mat}$ | $\Omega_\nu$ | $\sigma(8h^{-1}{\rm Mpc}, 0)$ | $\Gamma$ | $z_{\rm ion}$ | | |
|---|---|---|---|---|---|---|---|---|---|
| | | | | | | | low estimate | best guess | high estimate |
| Standard CDM | 0.5 | 1.0 | 1.0 | 0 | 1.41 | 0.45 | 28 | 46 | 69 |
| Tilted CDM, $n = 0.7$ | 0.5 | 0.7 | 1.0 | 0 | 0.71 | 0.31 | 6 | 10 | 16 |
| CDM, $h = 0.3$ | 0.3 | 1.0 | 1.0 | 0 | 0.64 | 0.23 | 8 | 13 | 20 |
| MDM, $\Omega_\nu = 0.15$ | 0.5 | 1.0 | 1.0 | 0.15 | 1.08 | 0.27 | 16 | 27 | 44 |
| MDM, $\Omega_\nu = 0.30$ | 0.5 | 1.0 | 1.0 | 0.3 | 0.93 | 0.19 | 7 | 13 | 22 |
| CDM, $\Omega_{\rm vac} = 0.7$ | 0.8 | 1.0 | 0.3 | 0 | 1.64 | 0.23 | 25 | 39 | 61 |
| Open CDM | 0.8 | 1.0 | 0.3 | 0 | 0.48 | 0.23 | 11 | 19 | 29 |

For the matter density contrast the suppression factor is accurately parametrised by (Carroll et al. 1992)

$$g = \frac{5}{2}\Omega_{\rm mat}\left[\Omega_{\rm mat}^{4/7} - \Omega_{\rm vac} + \left(1 + \frac{\Omega_{\rm mat}}{2}\right)\left(1 + \frac{\Omega_{\rm vac}}{70}\right)\right]^{-1}. \quad (16)$$

To use this result we need the COBE normalisation. A complete discussion along the lines of Bunn et al. (1994) has not been given, but the literature contains sufficient information that we can extract a reliable normalisation.

For the cosmological constant model there is no spatial curvature, so the Fourier expansion can still be used to define the spectrum $\delta_H^2(k)$. As in the standard CDM model we take it to be scale independent, though note that the most detailed comparison of this model with observations (Kofman et al. 1993) included some tilt. With this assumption, the COBE data have been fitted by Bunn & Sugiyama (1994); they find that the overall fit to COBE is quite poor (see also Sugiyama & Silk (1994)), and quote a best fit $Q_{\rm rms-PS} = 21.6\mu{\rm K}$. Unfortunately they do not give the corresponding value of $\delta_H$, but we have calculated it by making the reasonable assumption that the Sachs-Wolfe approximation is good for the quadrupole. Because the potential fluctuations are induced by $\delta\rho$ rather than $\delta\rho/\rho$, the matter fluctuations are then bigger by a factor $1/\Omega_{\rm mat}$. The correction due to the time dependent potential induced by the cosmological constant has been calculated by Kofman & Starobinsky (1985); for a given $\delta_H$, the expected quadrupole is about 9% bigger than in the critical case for the value $\Omega_{\rm mat} = 0.3$. This increase happens to coincide with the increase in the best fit value of $Q_{\rm rms-PS}$, so we conclude that the COBE normalisation of $\delta_H$ is larger just by the factor $1/0.3$. The central reionization redshift is therefore $z_{\rm ion} = 39$, the closest to the standard CDM value of all the models we have studied. Including the growth suppression factor $g = 0.78$ gives the other results in Table 1.

For the open universe model, there is no unique generalisation of the concept of scale-invariance in the spectrum on scales comparable to the curvature scale. This makes the COBE normalisation ambiguous. To break the ambiguity, we take the view that the scale dependence should be that given by the inflationary prediction using the conformal vacuum (Lyth & Stewart 1990).

Using the inflationary shape for the spectrum, Sugiyama & Silk (1994) have calculated the $C_l$. For $\Omega_{\rm mat} = 0.3$ they find that the shape mimics the critical density Sachs-Wolfe prediction with a spectral index $n \simeq 1.4$. As mentioned earlier, Górski et al. (1994) found that the COBE data can be fitted with $0.6 < n < 1.4$, so this value is marginally allowed, though as in the cosmological constant case the fit will be poor compared with the critical density case. We therefore normalise the open model by requiring that the predicted value of $C_9$ agrees with the value found in the fit of Górski et al. To determine this normalisation, we use values of $C_l$ given by Kamionkowski et al. (1994) (they actually normalise to the ten degree variance of the temperature anisotropy, but they also give the $C_l$). The resulting normalisation of $\delta_H$ is 0.49 times that for critical density, leading to a later reionization redshift $z_{\rm ion} \simeq 19$. Further, the suppression factor of $g = 0.45$ is stronger than for the cosmological constant model, leading to a much lower value for $\sigma(8h^{-1}{\rm Mpc}, 0)$.

Concerning whether these models fit the large scale structure data, we note that the observations must be reinterpreted. The shape is still given by $\Gamma$ from (11), but the observational value of $\sigma(8h^{-1}{\rm Mpc}, 0)$ increases. Peacock & Dodds (1994) give the scaling of their power spectrum with $\Omega$, and White et al. (1993) give the scaling of the cluster abundance. In combination, these give a target range

$$\sigma(8h^{-1}\,{\rm Mpc}, 0) = 0.9 \text{ to } 1.4 \quad (17)$$

in either open or cosmological constant models with $\Omega_{\rm mat} = 0.3$. We shall make a more detailed comparison of these models with observation elsewhere.

## 4 CONCLUSIONS

The most important aspects of the results that we have presented here are firstly that we have normalised all of the models to COBE, and secondly that we have taken account of baryonic corrections to the transfer functions. The normalisation to COBE is at the high end of the possibilities considered by TSB for standard CDM, but is considerably lower than that they considered for variant models. The in-clusion of baryons reduces the result for $z_{\rm ion}$ by about 20% in typical models. For $\Omega_{\rm mat} = 1$, we have presented a collection of observational data across a wide range of scales in a suitable form for extrapolation towards the scales relevant for reionization. It is clear that observations favour an epoch of reionization considerably later than that found in the standard CDM model.

We then examined a series of COBE normalised models. Most options favoured by the data give reionization around a redshift ten to twenty. However, the cosmological constant model, with its significantly higher normalisation of fluctuations, gave an answer considerably higher than the rest and almost as high as the standard CDM model. The other type of model capable of giving rather early reionization is an MDM model with a low choice for the density of the hot component; this arises via a combination of modest growth suppression and the model's natural tendency to explain the shape of the galaxy correlation function with a higher $\sigma(8h^{-1}{\rm Mpc}, 0)$ than other models.

The most important hurdle that models must surpass in connection with reionization is the Gunn–Peterson test (Gunn & Peterson 1965). For some variant CDM models, this could conceivably be tricky if the most pessimistic scenario is realised, but the question clearly cannot be addressed with existing understanding. Lower limits on the short scale perturbation spectrum from early object formation such as damped Lyman alpha systems (Mo & Miralde-Escude 1994; Kauffmann & Charlot 1994; Ma & Bertschinger 1994) promise to be much more secure.

The other role that reionization can play is in the damping of degree scale microwave background anisotropies (Vittorio & Silk 1984; Bond & Efstathiou 1984). This damping has been analysed recently by Sugiyama et al. (1993), and proves to be very sensitive to the reionization redshift. At $z_{\rm ion} \simeq 50$, towards the top end of the standard CDM predictions, the Doppler peak may be damped almost completely away. Even if $z_{\rm ion}$ is around 20, which appears more feasible, they find a suppression of over 30% which is certainly significant and seems likely in the cosmological constant model. By a redshift of 10, appropriate to several models, the effect is very small.

**ACKNOWLEDGEMENTS**

ARL is supported by the Royal Society and acknowledges the use of the Starlink computer system at the University of Sussex. We thank Ted Bunn, John Peacock, Douglas Scott, Andy Taylor, Peter Thomas, Martin White and Pedro Viana for discussions.

This paper has been produced using the Blackwell Scientific Publications LaTeX style file.